# The Cosmic Ray Helium and Carbon Nuclei
# Spectra Measured by Voyager 1 at Low Energies and Earth
# Based Measurements of these Nuclei up to 200 GeV nuc
# Concordance at High Energies with a Leaky Box Propagation Model

W.R. Webber[1] and P.R. Higbie[2]

1. New Mexico State University, Astronomy Department, Las Cruces, NM 88033, USA
2. New Mexico State University, Physics Department, Las Cruces, NM  88003, USA

## ABSTRACT

A comparison of the Helium and Carbon interstellar spectra measured at Voyager in the local interstellar medium leads to a different interpretation than a comparison of the Hydrogen to Helium spectra. This is because the He/C ratio is observed to increase rapidly with energy below ~40 MeV/nuc in contrast to an almost constant H/He ratio at these low energies. Both the He and C spectra that are observed at Voyager above 40 MeV/nuc and much higher energy spectra from the PAMELA measurements of these two components up to ~100 GeV/nuc can be accurately fit to within $\pm$ 10% assuming galactic propagation in a leaky box type of diffusion model in the galaxy with identical source spectra $\sim P^{-2.28}$ for He and C using a diffusion coefficient $\sim P^{0.50}$ above ~1 GV rigidity. These same exponents also fit the H spectrum from ~40 MeV to over 100 GeV. At low energies an excess of He relative to C is observed that would amount to about 20% of the modeled galactic component at 10 MeV/nuc.

## Introduction

The spectra of galactic cosmic ray (GCR) nuclei down to energies of a few MeV/nuc have been measured for the first time beyond the heliopause on V1 (Stone, et al., 2013). The study of these particles is being continuously updated and much of the data used in this article comes from Cummings, et al., 2014. In the 1[st] of a series of papers examining the implications of these new measurements (Webber and Higbie, 2015, hereafter referred to as paper 1), a comparison of the measured H and He nuclei spectra by Voyager in the energy range below ~300 MeV/nuc was made with the interstellar spectra that were calculated using a Leaky Box Model (LBM) for interstellar propagation. A feature of this comparison was the nearly constant H/He nuclei intensity ratio below a few hundred MeV/nuc seen in the data and was reproduced in the propagation calculations. The second significant feature of this H and He spectral comparison was the ability to fit these new low energy H and He spectra with existing higher energy H and He spectra from BESS (Sanuki, et al., 2000) and from PAMELA (Adriani, et al., 2013) on these nuclei up to >100 GeV/nuc, by using rigidity source spectra ~$P^{-2.28}$, with a nearly constant source exponent throughout this entire range from ~50 MeV/nuc up to ~100 GeV/nuc.

In this paper we make a detailed comparison between the Voyager measurements of He and C nuclei and the interstellar spectra calculated for these two nuclei. These nuclei are the most abundant of the GCR source nuclei with A/Z ~2. The measurements of both the spectra and charge ratios of these nuclei are therefore one of the most sensitive probes of the galactic propagation effects on the newly revealed low energy spectra observed by Voyager. Also He has a strong anomalous component at lower energies in the heliosheath but for C nuclei this anomalous component is very weak.

In the case of He and C, in contrast to H and He discussed in paper 1, the observed low energy He/C ratio is strongly dependent on energy/nuc whereas the H/He ratio is nearly constant at lower energies. In this paper this He/C ratio comparison is extended up to ~100 GeV/nuc and the He and C spectra and He/C ratio are calculated using the same galactic propagation model parameters as in paper 1.

**The Data**

The data for the He and C spectra are shown in Figure 1. This figure includes the new Voyager 1 He and C spectra from Cummings, et al., 2014, beyond the heliopause at low energies and also the higher energy He measurements from spacecraft near the Earth described in paper 1 plus the higher energy Carbon spectrum measurements from Engelmann, et al., 1990, and the new PAMELA measurements on Carbon extending up to ~100 GeV/nuc (Adriani, et al., 2014). The He intensities, plotted in Figure 1, are divided by 10 and the figure is a split figure with the Y-axis scale changing from left hand to right hand at 1 GeV/nuc, the same as paper 1.

Figure 2 shows the He/C intensity ratio obtained from the Voyager data at lower energies. At energies above ~500 MeV/nuc the He/C ratio in the figure is obtained from the He data (Adriani, 2001) and the C data (Adriani, et al., 2014) from PAMELA, both from the same time period.

The observed E/nuc dependence of the He/C ratio in Figure 2 is completely different from that of the H/He ratio described in paper 1, Figure 3. At energies above ~1.0 GeV/nuc the He/C ratio underline{decreases} slowly with increasing energy from a value ~40 at 1.0 GeV/nuc, and would reach the source ratio ~25 above a few hundred GeV/nuc as the path length, λ, approaches zero. This is opposite to the H/He ratio which underline{increases} from a value ~12 at a few hundred MeV/nuc to a constant source E/nuc ratio ~17.5 at energies above ~10 GeV/nuc (see Figure 3 of paper 1). At the energies below a few hundred MeV/nuc that are observed by Voyager, the H/He intensity ratio becomes nearly constant whereas the He/C intensity ratio suddenly begins a rapid increase at energies below ~40 MeV/nuc reaching values >100 at the lowest energies, a value which is over 4 times the source ratio of 25.

The solid black curves in Figures 1 and 2 are the intensities, dj/dE, of He and C respectively, from the LBM galactic propagation calculation for source rigidity spectra, dj/dP, with an exponent =-2.28, which is constant as a function of rigidity; along with a dependence of path length, λ, which is ~$P^{-0.5}$, also independent of rigidity, above 1-2 GV. These are the same source spectra and rigidity dependence for the mean path length, λ, used for the propagation of H and He nuclei in paper 1. This combination of parameters leads to rigidity (or E/nuc) spectra for both He and C which have exponents -(2.28+0.50) = -2.78 above ~100 GV. The calculated

E/nuc spectra of He and C nuclei in Figure 1 of this paper, and also Figures 1 and 2 of paper 1 for H and He nuclei, and the observed spectra of all of these three nuclei are indeed all approaching this exponent at energies above ~100 GeV/nuc (Adriani, et al., 2011, 2014).

**Diffusive Cosmic Ray Propagation in the Galaxy**

The galactic propagation model used in this paper is the identical LBM used in paper 1 (see also Webber and Higbie, 2009). The reader is referred to paper 1 and references therein for details of this model which was originally developed at SACLAY and used by Engelmann, et al., 1990, to compare with their data on Be-Ni nuclei measured on the HEAO-C satellite. Since the reacceleration term in the LBM is set =0, the key parameters affecting the spectral shape and intensities are; (1) the mean path length, $\lambda$, in g/cm$^2$ which is $\lambda = 26.5\ \beta R^{-0.5}$ above 1.0 GV and its rigidity dependence at lower rigidities which is important for this paper; and (2), at energies below ~100 MeV/nuc, the energy loss term which is mainly the ionization loss as the various nuclei traverse the interstellar matter.

The details of the uncertainties in $\lambda$ above 1.0 GV, which mainly depend on fitting the observed cosmic ray B/C ratio as a function of energy, are described in paper 1. The appropriate details of the cross sections, matter densities, etc., are also described in paper 1. In this model the path length distribution is a simple exponential at each mean path length.

In our comparison of the V1 data at low energies with the BESS and PAMELA H and He nuclei data at higher energies, as described in paper 1, the LBM calculations reproduced the observed intensities and spectra of both H and He nuclei using a source rigidity spectrum ~$P^{-2.28}$, independent of rigidity between about 0.5 and 200 GV along with a path length $\lambda=26.5\ \beta\ P^{-0.5}$ above ~1.0 GV. A break in the rigidity dependent path length is assumed to occur below 1.0 GV where the rigidity dependence of the diffusion coefficient is then assumed to change to $P^{1.0}$. This change in rigidity dependence of $\lambda$ is necessary to fit the observed H and He spectra at low energies, as well as the Voyager measured electron spectrum (Webber and Higbie, 2013). We illustrate the effects of this break for a value of $P_0 = 1.0$ in Figures 1 and 2 of this paper.

## Discussion – The Differences Between the Propagated and Observed Spectra for He and C Nuclei

First, from Figure 1 we see that the same LBM parameters used to obtain a fit to the H and He spectra provide an equally good fit to the C nuclei spectrum shown in Figure 1 over a broad energy range, which extends from the peak of the low energy C spectrum at ~50 MeV/nuc measured by V1, up to energies ~100 GeV/nuc measured by PAMELA (Adriani, et al., 2014), HEAO-C (Engelmann, et al., 1990) and others (see Seo, 2012). The calculations also provide a good fit to the higher energy He/C ratio data in Figure 2 where this ratio is seen to decrease from ~40 at 1 GeV/nuc to ~33 at 10 GeV/nuc as the ratio eventually approaches its source ratio of 25 as the path length → 0.

Collectively, all of the data above 40 MeV/nuc including the new Voyager data can be reasonably well fit with the propagation model used here with a $P_0 = 1.0$ GV for the path length dependence at low rigidities.

Below ~40 MeV/nuc the new V1 measurements show a rapid increase in the He/C ratio from a value of about 60 at 40 MeV/nuc to over 100 at 10 MeV/nuc. An increase in the He/C ratio is indeed predicted.

This predicted increase for a $P_0 = 1.0$ GV path length dependence used here does not match the observations, however. An examination of both the individual spectra of He and C and the He/C ratios in combination in Figures 1 and 2, suggests that two possibilities could be occurring to produce the increased He/C ratio that is observed at the lowest energies: (1) The ionization energy loss factor could be larger than we have used in any of the models. An increased rate of energy loss, as could occur from an increased ionized fraction of the interstellar medium, would reduce the C nuclei intensities relative to He, thus increasing the He/C ratio. In this paper we have used 15% for the ionized fraction. The results for an ionizing fraction of 30%, and for $P_0 = 1.0$ GV are insufficient to produce the observed increase in He/C ratio.

(2) The existence of a separate component of He in excess of the normal galactic He component that is predicted beyond the heliopause in a Leaky Box model. In this case, part of the intensity of He at the lower energies would be due to this component. This would increase the He/C ratio and be an "excess" of the He spectrum relative to the rapidly decreasing C spectrum.

In this scenario then the C spectrum at low energies has the normal predicted shape of the interstellar spectra of all cosmic ray primary nuclei in a Leaky Box model. The extra He component would need to have an intensity of about 20% of the total He intensity measured at 10 MeV/nuc to make the remaining LIS He spectrum at low energies "look" like that for propagated C and other heavier source nuclei. In order to pursue this scenario of an added He component at low energies the spectra from other less abundant heavier nuclei need to be studied with improved statistics. This project is underway and involves possible excesses of both H and He nuclei relative to the spectra of heavier nuclei from C to Fe. It also involves the possibility of a nonstandard Leaky Box model in which the path length distribution is no longer exponential but has a deficiency of short path lengths as could be expected in a near source model (e.g., Lezniak and Webber, 1979).

## Summary and Conclusions

In this paper we compare the new low energy He and C spectra that have been observed by Voyager 1 beyond the heliopause, along with available higher energy data up to and beyond 100 GeV/nuc, with calculations using a LBM diffusion model for galactic propagation. Again, as with the H and He spectral comparisons reported in paper 1, we find that a source rigidity spectrum $\sim P^{-2.28}$ for both nuclei and with a constant source exponent over the rigidity range 50 MeV/nuc to 100 GeV/nuc (-0.5 to 200 GV) along with a diffusion coefficient $\sim P^{-0.5}$ above 1.0 GV is consistent with the data on He and C nuclei at both low and high energies. This includes the peak in the spectra of He and C nuclei observed by V1 at between about 30-60 MeV/nuc. Below this energy, however, the observed C nuclei spectrum falls off more rapidly than that of He thus resulting in a rapidly increasing He/C ratio at lower energies. The propagation models do predict an intensity decrease of C nuclei relative to He because of the increased energy loss due to ionization which is $\sim (Z^2/A)$, but the observed increase in the He/C ratio is much greater than these predictions even for an ionized fraction of 30%.

We consider this spectral difference of He and C as possible evidence of a separate low energy component of He with an intensity 20% of the normal galactic cosmic ray spectrum He at 10 MeV/nuc.    In any case there also appears to be spectral differences between the primary charges He and C that are unaccounted for in the idealized LBM we have used.

**Acknowledgements:**  The authors appreciate the efforts of all the Voyager CRS team, Ed Stone, Alan Cummings, Nand Lal and Bryant Heikkila.  It is a pleasure to work with them as we try to understand and interpret the galactic cosmic ray spectrum which has previously been hidden from us by the solar modulation.  We also thank JPL for their continuing support of this program.

## References

Adriani, O., Barbanino, G.C. Bazilevskaya, G.A., et al., 2013, Ap.J., 765, 91

Adriani, O., Barbanino, G.C. Bazilevskaya, G.A., et al., 2014, Ap.J., 791, 93

Cummings, A.C., Stone, E.C., Heikkila, B., et al., 2014, AAS, Head Meeting #14, #201.01

Engelmann, J.J., Ferrando, P., Soutoul, A., et al., 1990, A and A, 233, 96-111

Lave, K.P., Wiedenbeck, M.E., Binns, W.B., et al., 2013, Ap.J., 770, 117

Lezniak, J.A. and Webber, W.R., 1979, Astrophys. and Space Sci., 63, 35-56

Sanuki, T., Matola, M., Matsumote, H., et al., 2000, Ap.J., 545, 1135

Seo, E.C., 2012, Centenary Symposium, Discovery of Cosmic Rays, Ed. J.F. Ormes, AIP, 1516, 181-185

Stone, E.C., Cummings, A.C., McDonald, F.B., et al., 2013, Science, 341, 150-153

Webber, W.R., McDonald, F.B. and Lukasiak, A., 2003, Ap.J., 599, 582-595

Webber, W.R. and Higbie, P.R., 2009, JGRA, 114, 2103

Webber, W.R. and Higbie, P.R., 2013, http://arXiv.org/abs/1308.6598

Webber, W.R. and Higbie, P.R., 2015, http://arXiv.org/abs/1503.01035

**Figure Captions**

**Figure 1:** The He and C spectra measured at Voyager beyond the heliopause (Stone, et al., 2013; Cumming, et al., 2014). This is a split figure. Below 1 GeV/nuc the intensities are given by the left hand vertical axis. Above 1 GeV/nuc the intensities are x $E^2$ (GeV). The BESS 97-98 measurements (Sanuki, et al., 2000), and PAMELA measurements above 1 GeV/nuc for He (Adriani, et al., 2013) are shown. For C nuclei, the measurements from HEAO-C (Engelmann, et al., 1990) and the new measurements from PAMELA (Adriani, et al., 2014) are also shown. The LBM propagation calculations are shown for a parameter $P_0 = 1.0$ GV as described in the text.

**Figure 2:** The He/C ratio as a function of energy/nuc. At energies below ~150 MeV/nuc the measurements are from Voyager 1 (Stone, et al., 2013; Cummings, et al., 2014). At 500 MeV/nuc and above the He/C ratio is from corresponding PAMELA measurements of both He (Adriani, et al., 2013) and C (Adriani, et al., 2014) up to ~100 GeV/nuc. The LBM propagation calculations are shown for a parameter $P_0 = 1.0$ GV and for a fraction of ionized H = 15% as described in the text.

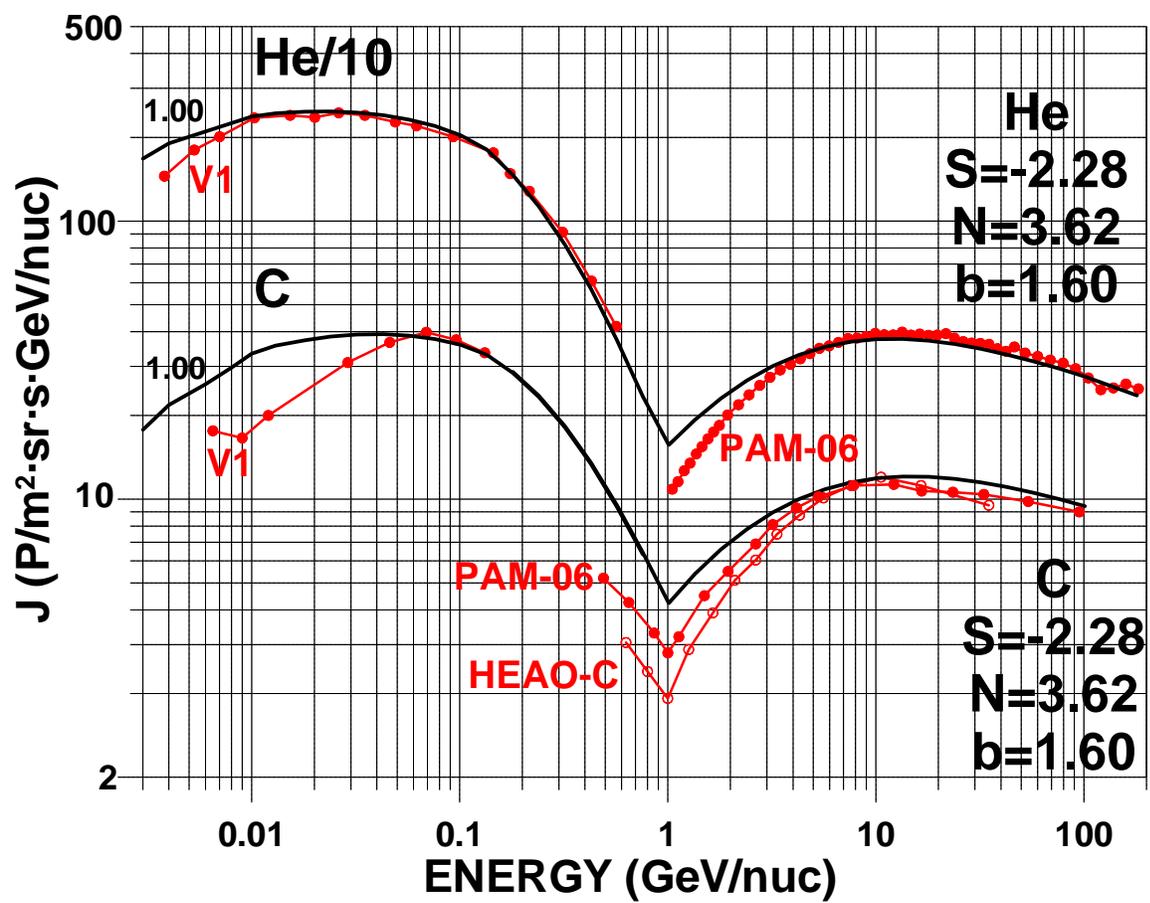

FIGURE 1

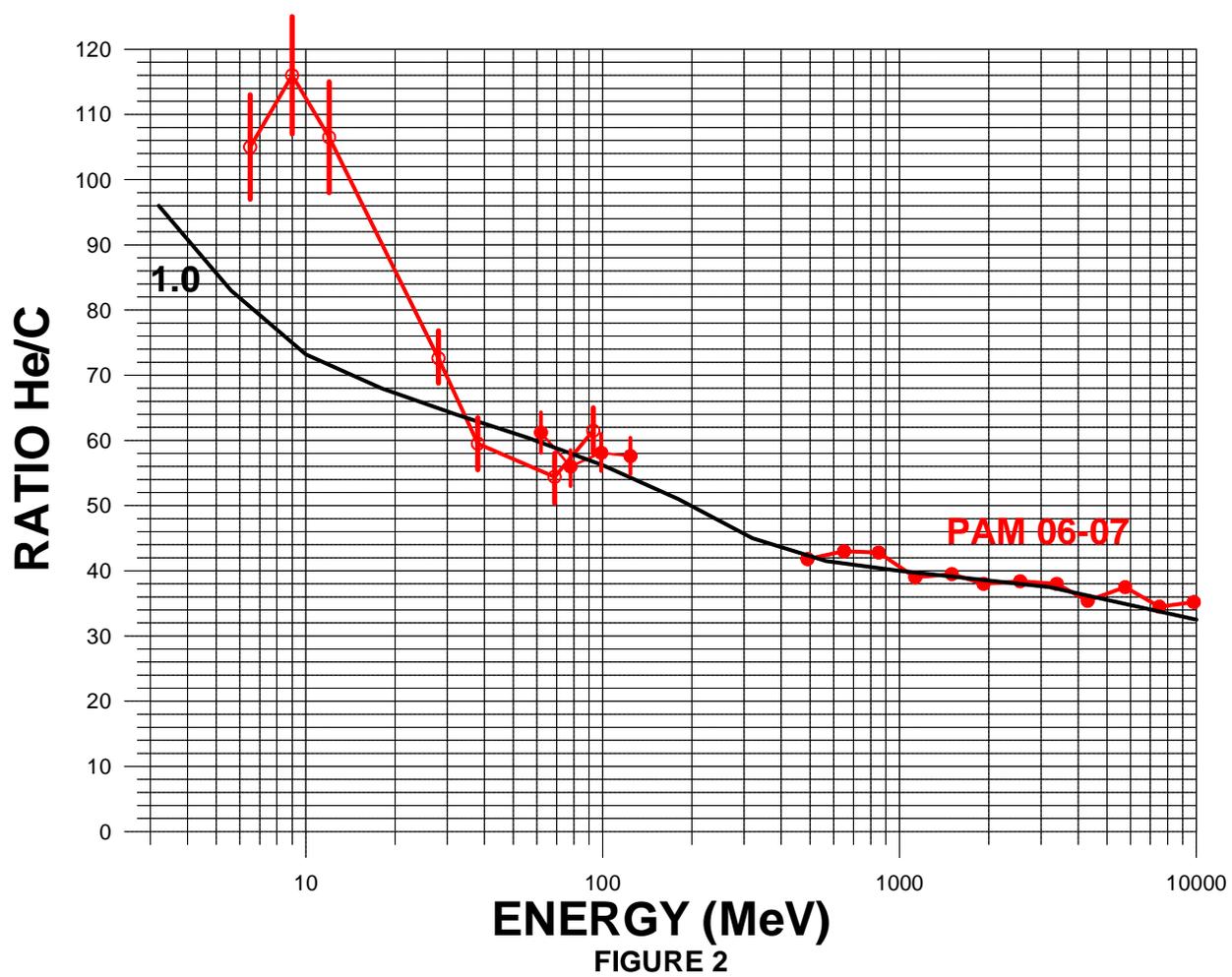

**FIGURE 2**